\documentclass[aps, prb, superscriptaddress, twocolumn, letterpaper, 10pt, amsfonts, amsmath, amssymb]{revtex4-2}
\usepackage{graphicx}
\usepackage{dcolumn}
\usepackage{bm}
\usepackage[pdfencoding=auto, psdextra, hidelinks]{hyperref}
\usepackage{physics}
\usepackage{float}
\usepackage{booktabs}
\usepackage{bbold}
\usepackage{nicefrac}
\usepackage{soul}
\usepackage{placeins}
\usepackage[normalem]{ulem}
\usepackage{xcolor}
\usepackage{lmodern}

\definecolor{edit_color}{rgb}{0.8, 0, 0}
\newcommand{\edit}[1]{{\color{black} #1}}

\definecolor{AMcolor}{rgb}{0.0, 0, 1.0}

\begin{document}
\title{Vortex nucleation barriers and stable fractional vortices near boundaries in multicomponent superconductors}

\date{\today}

\author{Andrea Maiani}
\affiliation{Center for Quantum Devices, Niels Bohr Institute, University of Copenhagen, DK-2100 Copenhagen, Denmark}

\author{Andrea Benfenati}
\affiliation{Department of Physics, The Royal Institute of Technology, Stockholm SE-10691, Sweden}

\author {Egor Babaev}
\affiliation{Department of Physics, The Royal Institute of Technology, Stockholm SE-10691, Sweden}

\begin{abstract}
The magnetization process of a superconductor is determined by the potential barrier for vortex nucleation and escape. In multicomponent superconductors, fractional vortices with a winding in the phase of only one of the components can be stable topological solitons that carry a fraction of the flux quantum. While the formation of such objects in the bulk costs logarithmically or linearly divergent energy, these objects were shown to be stable near samples' boundaries in the two-component London model. Therefore, the conventional Bean-Livingston picture of magnetic flux entry does not apply to these superconductors, since the entry process can involve fractionalization of a vortex. 
In this paper, we address the nonlinear problem of determining the potential barrier for fluxoid penetration in a multicomponent superconductor, including the effects of various intercomponent couplings, by using the recently developed gauged string method. The method allows numerically exact (i.e., convergent) calculation of a sphaleron configuration in a gauge theory and thus the height of the nucleation barrier. We show how the fractionalized nucleation processes result in multiple sphalerons and intermediate states due to the complex shape of the energy landscape of multicomponent superconductors.
\end{abstract}

\maketitle

\section{Introduction}
Most superconductive systems of current interest have multiple components. That include multi-band superconductors with $s$-wave pairing, see e.g.~\cite{Suhl1959,moskalenko1959superconductivity,kamihara2008iron,nagamatsu2001superconductivity,kreisel2020remarkable,Svistunov2015}, superconductors with broken time-reversal symmetry, which can be singlet $s$-wave superconductors~\cite{grinenko2017superconductivity,grinenko2018emerging,grinenko2021bosonic,kivelson2020proposal}, and superconductors materials with unconventional pairing~\cite{sigrist1991phenomenological}. When a superconductor has multiple components, the vortex excitations have a composite character: they are bound states of fractional-flux vortices, which, individually, have logarithmically divergent energy due to inter-component electromagnetic coupling, or linearly divergent density due to phase-difference-locking coupling, and therefore, form finite-energy bound states~\cite{Babaev2002}. This affects the magnetic properties of such materials since, to enter a superconductor, a vortex needs to overcome a potential barrier. 

In the Bean-Livingston picture, the origin of the surface barrier for a vortex is described within the London model, as the effect of the competition between the attractive force of a mirrored image-vortex with opposite vorticity, and the repulsive force with the surface current induced by the external magnetic field~\cite{Bean1964a}. However, in general, the vortex position in the maximum energy configuration can be very close to the boundary, preventing the use of the superposition principle that holds only for linear theories. Therefore, a general description requires the solution of the full nonlinear problem, e.g. in a Ginzburg-Landau model. 

These energy barriers correspond to infinite-dimensional saddle points of the free energy landscape of the theory, and are called sphalerons, which are often associated with a transition between topologically distinct minima~\cite{manton2004topological, rubakov2009classical, Nastase2019}. These saddle points in the configuration space can be better understood as the local maximum of the minimum energy path of the nucleation process. That is defined as a path in the configuration space such that it crosses the minimum in the cotangent space of the path point by point.

In two-dimensional single-component systems, the number of processes featuring sphalerons is limited because the intervortex interaction is always repulsive, while the low dimensionality excludes processes involving complicated real-space topology changes like vortex reconnections. Indeed, sphalerons appear only in changes of the vortex lattice or in processes where vortices interact with the boundaries of the sample, like in the nucleation process.

When a second complex field comes into play, the energy landscape becomes more complicated and many new configurations, both stable states and sphalerons, are possible. In the nucleation process of a composite vortex, the vortices in the two components can experience attraction and repulsion to the boundary at different distances. The presence of different zones of attraction and repulsion can lead to the presence of barriers and metastable intermediate states. Vortex entry in a multicomponent superconductor is a much more complex process that, under certain conditions, involves the formation of a stable fractional vortex near the boundary. Stable fractional vortices in two-component systems were demonstrated both within the London model~\cite{silaev2011stable} and numerically in the Ginzburg-Landau model~\cite{agterberg2014microscopic}.

While many works, both analytical and numerical, focused on the calculation of the nucleation fields, that is the magnetic field needed to suppress the barrier~\cite{de1965vortex, catelani2008temperature,  transtrum2011superheating, Liarte2017, Pack2020}, the calculation of the barrier height and the sphaleron configuration is very complicated and, until recently, there has been no analytical nor numerical fully-controllable approach to address this problem in gauge theories such as Ginzburg-Landau models. One of the many methods to compute the minimum energy path for other models is the string method~\cite{E2002a, vanden2010transition}, originally designed to compute saddle points in molecular dynamics problems. In Ref.~\onlinecite{Benfenati2020} it has been proposed a generalization of the string method for classical gauge theories that allows calculating, in a numerically controllable way, vortex nucleation processes in a single-component Ginzburg-Landau model.

In this paper, we generalize to the multicomponent case the gauged string method presented in Ref.~\onlinecite{Benfenati2020} that allows studying, extensively and free of approximations, the process of vortex nucleation and to calculate vortex nucleation barriers. The main objective is to study how nucleation is affected by the presence of a second component, looking for fractional vortex nucleation and edge pinning. Secondarily, this allowed the testing of the gauged string method for a system with more complicated gauge symmetry.

\section{Model}
\edit{
Multicomponent Ginzburg-Landau theories have beeen microscopically derived for a variety of models~\cite{tilley1964ginzburg,agterberg2014microscopic, wagner2021microscopic,  chichinadze2020nematic, maiti2013s+, garaud2017microscopically,
 Silaev.Babaev:12, Krogh2018derivation}. 
Here, we consider a two-component model in which the order parameter $\Psi$ is represented by two complex scalar fields $(\psi_1, \psi_2)$. The two component fields are coupled by some direct interaction $V_\mathrm{int}$, and to the electromagnetic field described by the gauge potential $\vb A$.}
The system has a free energy $F\qty[\vb A, \psi_1, \psi_2] = \int \mathcal{F} \dd[3]{\vb{r}} $ where the free energy density functional reads, in natural units, 
\begin{widetext}
\begin{equation}
\begin{split}
    \mathcal{F}[\vb A, \psi_1, \psi_2] &= \sum_{\alpha=1,2} \qty[ \sum_{k=x,y} \frac{\abs{D_{k\alpha}\psi_\alpha}^2}{2 m_{\alpha k}} + \frac{b_\alpha}{2} \qty( \frac{a_\alpha}{b_\alpha} + \abs{\psi_\alpha}^2)^2] + \frac{1}{2}\qty(\curl \vb A - \vb H)^2 + V_{\mathrm{int}}\qty(\psi_1, \psi_2)\,,\\
    V_{\mathrm{int}}\qty(\psi_1, \psi_2) &= \frac{\eta}{2}\qty(\psi_1 \psi_2^* + \mathrm{c.c.}) +  \frac{\gamma}{2} \abs{\psi_1}^2 \abs{\psi_2}^2\,.
\end{split}
\label{eq:ni-mgl}
\end{equation}
\end{widetext}
Here the index $k$ labels the axes of the coordinate system and the index $\alpha$ denotes the superconducting components, the vector field $\vb A$ is the gauge field while $\psi_\alpha = \abs{\psi_\alpha} e^{i\theta_\alpha}$ are the two matter fields, and $\vb H$ is the external magnetic field. The parameters $m_{\alpha k}$ are the masses of the two fields in the reference system. \edit{The operator $D_{k\alpha} = -i\partial_k - q_\alpha A_k$ is the covariant derivative where $q_\alpha$ is the coupling constant of the component to the gauge field that parameterize the magnetic field penetration depth. For the direct interactions term, we only consider the linear Josephson coupling, controlled by the parameters $\eta$, and the density coupling $\gamma$. Higher gradient terms, additional fields, and more general interactions, including the effects of time-reversal symmetry breaking and nematicity in the interaction potential can be straightforwardly included.
}

The vortex topological invariants of the model are represented by the two winding numbers defined as:
\begin{equation}
    N_i = \int_{\partial \Omega} \grad \theta_i \cdot \dd{l}.
\end{equation} 
If we neglect the direct interaction term $V_{\mathrm{int}}\qty(\psi_1, \psi_2)$ the system described by the functional in Eq. \ref{eq:ni-mgl} models two charged superfluids, which only interact through the gauge field. In the absence of phase-locking or phase separation, the system has $U(1)\cross U(1)$ gauge symmetry. 

\edit{
 The magnetic flux carried by a vortex line, in the model that we consider,  is $\Phi = \sum_\alpha N_\alpha \Phi_\alpha$ with $\Phi_\alpha = \frac{2\pi q_\alpha \bar{\psi}_\alpha^2}{\sum_\beta q_\beta^2 \bar{\psi}_\beta^2}$ where $    \overline{\psi}_\alpha =\sqrt{\frac{-a_\alpha}{b_\alpha}} $ is the bulk equilibrium value of each order parameter, and $N_\alpha$ is an integer.
 
In conventional superconductors, $q_1=q_2=q$ and therefore it simplifies to 
\begin{equation}
    \Phi = \Phi_0 \left[ N_1 \frac{\overline{\psi}^2_1}{\overline{\psi}^2_1 + \overline{\psi}^2_2} + N_2 \frac{\overline{\psi}^2_2}{\overline{\psi}^2_1 + \overline{\psi}^2_2} \right]\,,
\end{equation}
where $\Phi_0=2\pi/q$ is the magnetic flux quantum. 

The elementary excitations in the model are \textit{fractional vortices}, which have a nonzero winding only in one phase~\cite{Babaev2002}.  Fractional vortices are thermodynamically unstable in the bulk~\cite{Babaev2002} but they can be stabilized near the boundary of a sample by the interaction with surface currents~\cite{silaev2011stable}. Only when the vortex carries an integer number of flux quanta, i.e. $N_1=N_2$, then the topological excitation is given the name of \textit{composite vortex} and it is also stable in the bulk of the superconductor.

Another case, that we consider, is a system  with condensates having different charges, that we parametrize as $q_i = g_i q$. With the simplifying assumption $\overline{\psi}_1 = \overline{\psi}_2$, the flux carried by a vortex is 
\begin{equation}
    \Phi = \Phi_0 \left[ \frac{g_1 N_1 + g_2 N_2}{g_1^2 + g_2^2} \right]\,.
    \label{eq:disparity_quantization}
\end{equation}

Also in this case, the only stable soliton in the bulk is the one having an integer number of flux quanta, which are composite vortices with a $(N_1, N_2) = (g_1, g_2)$ structure. We show later that, when the topological excitations do not satisfy the integer quantization of flux, they can exist only near the system's boundaries, because of a stabilization mechanism similar to the one for fractional vortices in conventional two-component superconductors.
}

\edit{
The interaction of vortices between themselves and the boundary are determined by the different length scales that are present in the model. In a isotropic system without direct interaction, i.e. $m_{\alpha x} = m_{\alpha y}$ and $V_\mathrm{int} = 0$, we can identify the magnetic field penetration depth as $\lambda = \qty(\sum_\alpha \lambda_\alpha^{-2})^{-1/2}$ where $\lambda_\alpha = \sqrt{\frac{m_\alpha}{q_\alpha^2\overline{\psi}_\alpha^2}}$, and the components coherence lengths as $\xi_\alpha =  \frac{1}{2 \sqrt{-m_\alpha a_\alpha}}$. If $\xi_1,\xi_2<\lambda$ the system behave as a type-2 superconductor, while if $\xi_1,\xi_2>\lambda$ it behaves as a type-1. 
Finally, the regime where $\min(\xi_1,\xi_2)<\lambda<\max(\xi_1,\xi_2)$ is termed type-1.5 superconductivity~\cite{babaev2005semi, Silaev.Babaev:11, Silaev.Babaev:12, Moshchalkov2009, Babaev2017, Svistunov2015, Babaev.Carlstrom:10, carlstrom2011type}.  In this regime, the interaction between vortices are long-range attractive and short-range repulsive. When an anisotropic mass tensor is introduced the length scales become direction-dependent increasing the number of possible regimes \cite{winyard2019hierarchies}. }

In models with direct intercomponent interactions, i.e. $V_\mathrm{int}\neq 0 $, it is not possible to associate separate coherence length to each component as the coherence lengths describe the decay of linear combinations of the matter fields~\cite{babaev2010type, carlstrom2011type, Silaev.Babaev:11}. In this scenario, is more convenient to speak of a \textit{large-core component} and a \textit{small-core component} based on the typical dimension of the fractional vortex cores, which we determine numerically.

Here, other types of solitons can be present. For example, two well separated fractional vortices can stick together. These objects can be described as skyrmions of the \edit{pseudospin vector field $\vb S = \frac{\Psi^\dag \vb*{\sigma}\Psi}{\Psi^\dag\Psi}$ where  $\Psi = (\psi_1, \psi_2)^T$ where and $\vb*{\sigma}$ are the Pauli matrices vector~\cite{Babaev.Faddeev.ea:02,agterberg2014microscopic}.}

An extremely useful tool to study the formation of solitons is the concept of minimum energy paths. A generic path in the configuration space of the system is a function $\vb q(s) = \qty(\vb A(s),\, \psi_1(s),\, \psi_2(s))$, where the transition coordinate $s\in[0, 1]$ has been introduced. A minimum energy path can be defined in a variational formulation as 
\begin{equation}
    \vb{q}_\mathrm{MEP}(s) = \arg \min F_\bot [\vb q(s)]\,, \quad \forall s\, \in [0,1]\,,
\end{equation}
where $F_\bot$ is that the free energy functional $F$ restricted to the cotangent space of the trajectory in the point $\vb{q}_\mathrm{MEP}(s)$~\cite{Samanta2013}.
Once the minimum energy path $\vb{q}_\mathrm{MEP}(s)$ is obtained, it is possible to compute the quantities of interest, like the free energy $F$ or the winding numbers $N_i$, and track their evolution along the path. The evolution along $s$ can be considered a pseudodynamics, which mimics some features of the real time dynamics simulated by time-dependent models. \edit{
We calculate minimum energy paths in the two-component Ginzburg-Landau model through the gauged string method introduced in Ref.~\onlinecite{Benfenati2020}. The details on the numerical implementation of this method can be found in the supplementary material of Ref.~\onlinecite{Benfenati2020}. For the reparametrization step of the method, we used the Euclidean distance calculated for the  observable magnetic field $\vb B = \curl \vb A$, and currents $j_{\alpha k} = \psi_\alpha^* \frac{D_{k \alpha} }{2m_{\alpha k}} \psi_\alpha + \mathrm{c.c.}$.
}

In this work, we mainly study the minimum energy paths of the single vortex nucleation process. The energy of the principal sphaleron, the most important barrier in the process, is just the maximum of free energy along the minimum energy path $F_\mathrm{S}=\max_s F(s)$. We define $F_\mathrm{M}$ as the energy of the Meissner state, which is a state with no vortices in the bulk. When studying the nucleation of the first vortex in the sample, the nucleation barrier is therefore defined as $\Delta F_\mathrm{n} = F_\mathrm{S} - F_\mathrm{M}$.

When looking at composite nucleation processes, the final point is a configuration when the vortex is in the center of the system. We denote the energy of this state as $F_\mathrm{C}$. If no surface bond states are present, the escape energy of the vortex is simply given by $\Delta F_\mathrm{e}=F_\mathrm{S}-F_\mathrm{C}$. When considering fractional vortices, the most stable position is not in the center of the sample but usually near the surface, where the vortex forms a bond with surface currents. We label the free energy of this configuration by $F_\mathrm{B}$. 

To characterize the surface-bonded fractional state, we are interested in evaluating the strength of the bond to the surface, by comparing the energy of the system with the same quantity when the vortex is in the bulk. However, the state with the vortex in the center of the bulk is not a stable state of the system but rather an unstable equilibrium point. For this reason, a minimum energy path cannot, in principle, be calculated. \edit{To circumvent this problem, we remove a single point from the superconductive domain to artificially introduce an extremely shallow local minimum, with negligible effects on the total energy of the field configuration. To minimize the perturbation of the free energy we choose point $(0,0)$ that is the farthest from the boundaries. We denote the energy of the configuration with the vortex pinned in the exact center of the domain by $F_\mathrm{C}$ also in this case}. 

\begin{figure}[!h]
    \centering
    \includegraphics[width=\columnwidth]{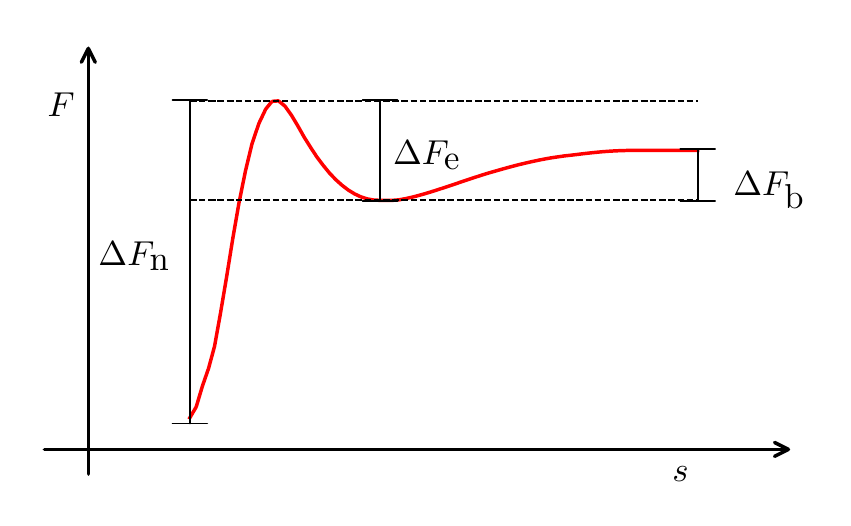}
    \caption{
    Example of a minimum energy path of the nucleation of a fractional vortex with the endpoint in the center of the superconductor domain. The picture shows the definition of nucleation barrier $\Delta F_\mathrm{n}$, escape barrier $\Delta F_\mathrm{e}$, and bonding energy $\Delta F_\mathrm{b}$. In case of absence of surface bonded vortex state $\Delta F_\mathrm{b} = 0$. To pin the vortex in the center of the domain the origin $(x,y)=(0,0)$ has been removed from the superconductive domain. This is necessary since the final configuration is not a minimum of the system. With this trick, the vortex is within a local minimum with negligible effects on the total energy. 
}
    \label{fig:minimum energy path_example}
\end{figure}

For the fractional nucleation curve, the escape barrier is then defined as the energy necessary to expel the vortex starting from the local minimum, $\Delta F_\mathrm{e} = F_\mathrm{S} - F_\mathrm{B}$,  while the bonding energy $\Delta F_\mathrm{b} = F_\mathrm{C} - F_\mathrm{B}$ is the energy required to move the vortex from the edge toward a region deep in the bulk. All these quantities are more clearly shown in Fig.~\ref{fig:minimum energy path_example}.

\edit{
For the simulations, we consider a system with computational domain $\Omega = [-L/2,+L/2]^{2}$ with $L=12$. We build the initial guesses for the component fields by moving the singularities from a point outside of the domain to the center. Notice that, the initial guess for the endpoint configurations at $s=0$ and $s=1$ are topologically equivalent to the ones in the converged minimum energy path as their motion follows a simple gradient descent.
}

\section{Results}
\edit{
The most general two-component Ginzburg-Landau model has a huge parameter space. For this reason, we show the results for some representative systems. For each series of simulations, we fix all parameters according to Tab.~\ref{tab:parameters} except for one or a pair. First, we focus on the case where the intercomponent direct interaction is absent, i.e. $V_\mathrm{int}=0$.
}

\begin{table}[h]
    \centering
\begin{tabular}{@{}cccccc|ccccc@{}}
\toprule
         & $a_1$ & $b_1$ & $m_{1x}$ & $m_{1y}$ & $q_1$ & $a_2$ & $b_2$ & $m_{2x}$ & $m_{2y}$ & $q_2$ \\ \midrule
System E & -1    & 1     & 1           & 1           & -1    & -1    & 1     & 1           & 1           & -1    \\ 
System U & -1    & 1     & 1           & 1           & -1    & -1    & 1     & 2.5         & 2.5         & -1   \\ 
System H & -0.5    & 0.5     & 0.5           & 0.5           & -1    & -0.5    & 0.5     & 4         & 4         & -1   \\ 
System D & -1    & 1     & 2         & 2         & -2    & -1    & 1     & 1         & 1         & -1    \\ \bottomrule
\end{tabular}
    \caption{Parameters used in the simulations. Since the model has many independent parameters, for each series of simulations we fix all the parameters as in this table except for one or a pair which are mentioned.}
    \label{tab:parameters}
\end{table}

\edit{
The characteristic length scales for the systems considered are shown in Tab.~\ref{tab:lengthscales}. We examine the nucleation in various systems: System E features two perfectly similar complex scalar fields in a type-2 regime, System U introduces different length scales but still belonging to the type-2 class. System H belongs to the type-1.5 class, while System D introduces condensates with different charges. In Systems U, H, and D, the two component fields have a different set of parameters and thus we identify $\psi_1$ as the large-core component while $\psi_2$ is the small-core one. 
}
\begin{table}[h]
    \centering
\begin{tabular}{@{}cccccccc@{}}
\toprule
 & $\xi_1$ & $\xi_2$ & $\lambda$ & $\bar{\psi}_1$ & $\bar{\psi}_2$ & $H_\mathrm{c}$ & $H_\mathrm{c1}$ \\
\midrule
System E & 0.50 & 0.50 & 0.71 & 1.00 & 1.00 & 1.41 & 1.16 \\ 
System U & 0.50 & 0.32 & 0.85 & 1.00 & 1.00 & 1.41 & 0.93 \\
System H & 1.00 & 0.35 & 0.67 & 1.00 & 1.00 & 1.00 & -\\
System D & 0.35 & 0.50 & 0.58 & 1.00 & 1.00 & 1.41 & - \\ 
\bottomrule
\end{tabular}
    \caption{
    \edit{The table reports the characteristic parameters of the system described in the paper. Component coherence length $\xi_i$ and magnetic penetration depth $\lambda$, equilibrium density $\bar{\psi}_i$,  thermodynamic critical field $H_\mathrm{c}$. The first critical field $H_\mathrm{c1}$ is defined as the external field that guarantee $F_M(H_\mathrm{c1}) = F_C(H_\mathrm{c1})$. Notice that the composite vortex considered for System D has a $(2, 1)$ structure.}
    }
    \label{tab:lengthscales}
\end{table}

\subsection{Surface bonded fractional vortices in type-2 systems}
The first case we address is the vortex nucleation processes in isotropic two-components type-2 systems.

\begin{figure}[!h]
    \centering
    \includegraphics[width=\columnwidth]{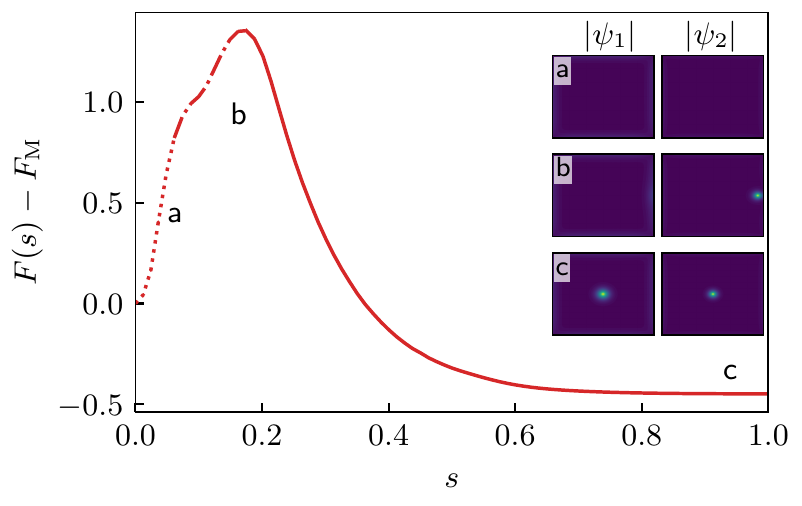}
    \caption{Minimum energy path of the nucleation process for a composite vortex in System U \edit{with $H=1.0$}. The minimum energy path shows an example of fractionalized nucleation: the nucleation process happens in two steps, with the small-core vortex nucleating first, followed by the large-core one. To report the nucleation steps, we use the following scheme: the dotted line signals that the systems has zero global winding number in both the components $(N_1, N_2) = (0,0)$, dash-dotted line means that only one component feature a non-zero winding number $(N_1, N_2) = (0,1)$, finally continuous line means both the components have a non-zero winding $(N_1, N_2) = (1,1)$.
    The insets show the matter fields configuration in three points along the minimum energy path. 
    }
    \label{fig:SysU_composite}
\end{figure}

In the case of a superconductor with two matter fields, having different characteristic lengths, like System U, we observe a \textit{fractionalized nucleation}. This means that the optimal nucleation process is, in general, split into two steps as the fractional vortices in each component enter the bulk in different stages. This is qualitatively in agreement with the earlier studies using the London model~\cite{silaev2011stable}. This behavior can be observed by computing the winding number for each field along the path $N_i(s)$ and comparing this curve with the free energy $F(s)$, as shown in Fig.~\ref{fig:SysU_composite}. Indeed, the small-core vortex nucleates first, followed by a vortex in the large-core component. The two fractional vortices are bound together in a composite vortex which then, pushed by the repulsion of surface currents, moves towards the center of the domain.

\begin{figure}[!h]
    \centering
    \includegraphics[width=\columnwidth]{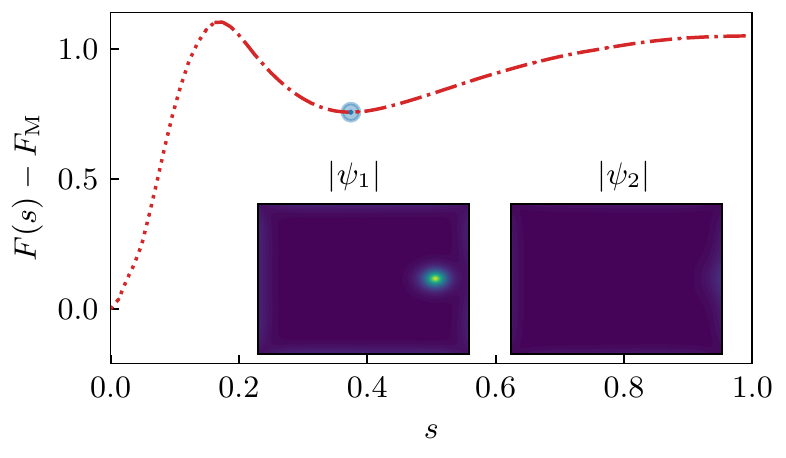}
    \caption{Free energy and winding numbers computed along the minimum energy path of the nucleation process for a fractional vortex with winding in the first component \edit{in System U at $H=1.0$}. If a vortex has non-zero winding only in the large-core component, it remains bonded to the surface. The curves flatten for $s\to 1$ because of the interaction of the vortex with edges in the finite-size system. 
    To report the nucleation steps, we use the following scheme: the dotted line indicates that the systems' winding numbers are $(N_1, N_2) = (0,0)$, while the dash-dotted line means $(N_1, N_2) = (1,0)$.
    }
    \label{fig:SysU_fractional}
\end{figure}

If, instead, we consider the nucleation of the fractional vortex with nonzero winding in the large-core component only, we see that a surface bonded state can exist in a region of the parameters space. The minimum energy path of the fractional vortex nucleation of this kind is shown in Fig.~\ref{fig:SysU_fractional}.

\begin{figure}[!h]
    \centering
    \includegraphics[width=\columnwidth]{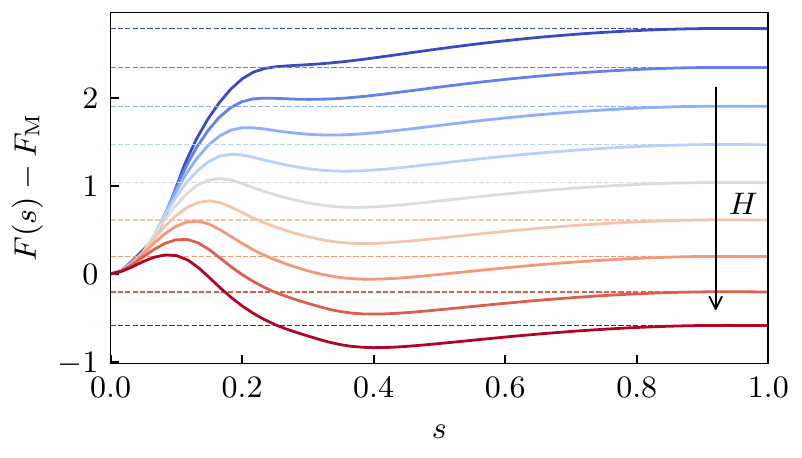}
    \caption{\edit{Stabilization of the surface fractional bonded state in System U, as the external magnetic field is increased from $H=0.6$ (blue line) to $H=1.45$ (red line)}. The solid lines represent the minimum energy path of fractional vortex nucleation. The dashed lines, representing the free energy of the system with the fractional vortex pinned in the center of the system, are included as a visual aid. $F_{\textrm{M}}$ indicates the free energy of the Meissner state.}
    \label{fig:fractional_formation}
\end{figure}

By studying the changes in the minimum energy path for fractional vortex nucleations, as a function of the external magnetic field, we verify that the stabilization mechanism occurs due to the interaction with the surface Meissner currents. The fractional vortex is attracted by the surface of the domain and forms a bond. As shown in Fig.~\ref{fig:fractional_formation}, an increase of the external magnetic field makes the minimum deeper.

It is possible to analyze the phase space of the system by extracting the free energy barriers $\Delta F_\mathrm{n}$, $\Delta F_\mathrm{e}$, $\Delta F_\mathrm{b}$ as a function of the external field. In this way, we can identify a region where fractional vortices are stabilized by surface currents. In the considered regime, the bonding energy $\Delta F_\mathrm{b}$ is approximately constant with respect to the magnetic field, meaning that the stability of the surface-bonded fractional state is weakly affected by the external magnetic field $H$, in the given field range. Fractional surface-bonded vortices can exist in the region between $H=0.7$ and $H=1.4$ while higher fields cause the nucleation of composite vortices. In this zone, the bonding energy is constant, while the escape barrier increases with the magnetic field intensity. We notice how, for fields higher than $H=1.2$, the escape barrier exceeds the nucleation one, meaning that a fractional vortex is more stable than the Meissner state. This means that, for fields in the region starting from $H=1.2$ to $H=1.4$, a surface bonded fractional vortex phase is expected to be the equilibrium state of the system. This is consistent with the magnetization simulations of a slightly different model~\cite{agterberg2014microscopic}.
    
\begin{figure}[!h]
    \centering
    \includegraphics[width=\columnwidth]{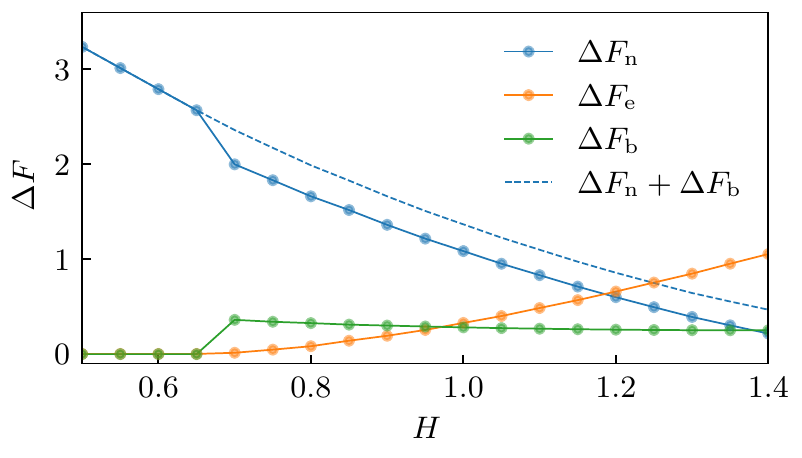}
    \caption{Free energy barriers for a fractional vortex in System U as a function of the external field $H$. At $H=0.7$, the surface fractional vortex states become metastable, as the escape barrier becomes nonzero. Fractional surface-bond vortices can exist in the region between $H=0.7$ and $H=1.4$, while higher fields cause the nucleation of composite vortices. In this zone, the bonding energy is constant while the escape barrier increase with the magnetic field intensity. Notice how, for fields higher than $H=1.2$, the escape barrier exceeds the nucleation one, meaning that a fractional vortex is more stable than the Meissner state. The bonding energy is approximately constant with respect to the magnetic field.}
    \label{fig:fractional_all}
\end{figure}

\begin{figure}[!h]
    \centering
    \includegraphics[width=\columnwidth]{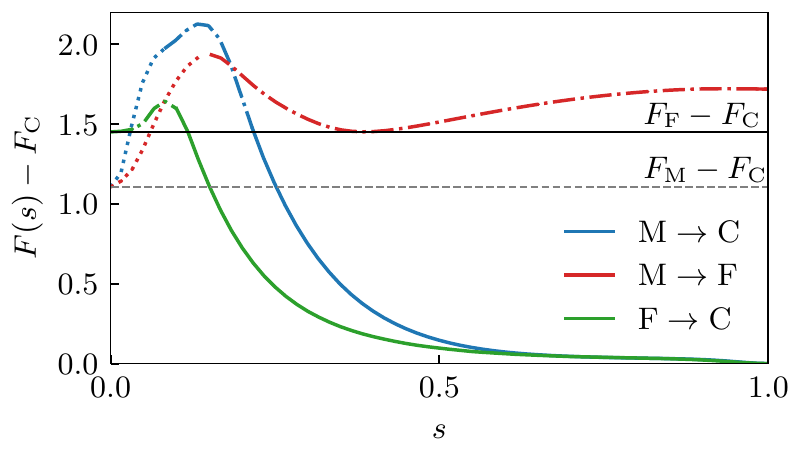}
    \caption{Comparison of nucleation minimum energy paths of a composite vortex ($\mathrm{C}$), a fractional vortex ($\mathrm{F}$) and the nucleation of a second component vortex starting from a fractional one ($\mathrm{F}\to\mathrm{C}$), in System U at $H=1.1$. 
    The black line is the energy of the system with a fractional vortex with respect to the composite vortex free energy $F_{\textrm{C}}$ and the gray line is the energy of the Meissner state with respect $F_{\textrm{C}}$.
    Notice how the nucleation of a fractional vortex (red line) has a barrier that is slightly smaller than the one of the composite vortex (blue line).
    Moreover, the nucleation process for a vortex in the second component starting from a fractional vortex bond to the surface has a very small barrier. This suggests that, in the given example, the fractional surface vortices are unstable to small fluctuations.
    Also in this case, the dotted line indicates the absence of winding in both components, the dashed-dotted line signals the presence of a fractional vortex, while the continuous line means that there is a composite vortex in the system. }
    \label{fig:fractional_comparison}
\end{figure}

It can be seen in Fig.~\ref{fig:fractional_comparison}, for the considered parameters, that, if the system has a vortex in the large-core component trapped near the surface, the activation energy for the nucleation of a vortex in the small-core component is usually low. The nucleation of a vortex in the other component frees the vortex from binding to the surface, as the composite vortex is repelled by the surface currents. This limits the lifetime of the surface-bonded state.

\subsection{Anisotropy effects}

The introduction of anisotropies in the model affects the nucleation process since it induces geometrical deformations in the vortices' cores. One way of introducing anisotropies in the model is to alter the ratio between the mass parameter of the fields. In the following, we consider the ratios $m_{2x}/m_{2y}=10$, and $m_{2y}/m_{2x}=10$. In all the simulations discussed in the paper, the vortex enters along the $x$ direction. The two cases considered here can, indeed, be seen as rotations of the same system or equivalently nucleation processes at two different sides.

\begin{figure}[!h]
    \centering
    \includegraphics[width=\columnwidth]{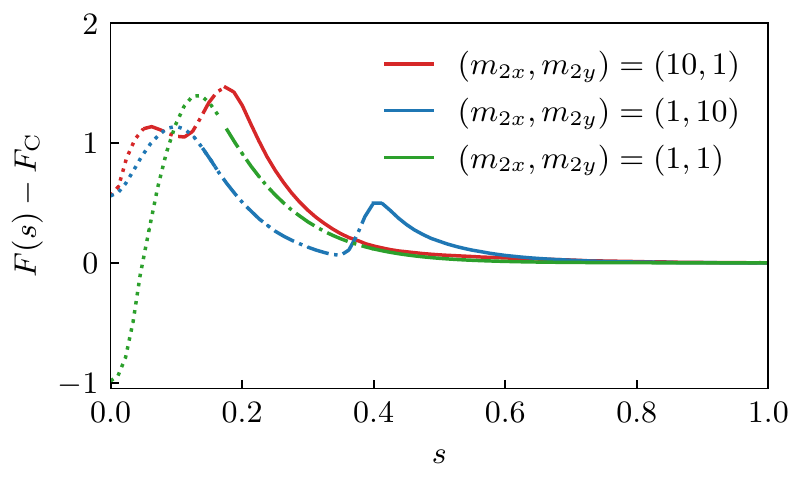}
    \caption{Minimum energy path of a composite vortex nucleation in a two-component superconductor with spatial anisotropy in one of the two components. The simulations show nucleation processes occurring in System E with various mass ratios at $H=1.0$. In these simulations, we study the vortex entry from a surface aligned with the $y$ direction. In the case where the heavy direction is parallel to the surface, the bond state energy is enhanced by the vortex core deformation. 
    To report the nucleation steps, we use the following scheme described in Fig.~\ref{fig:SysU_composite}
    }
    \label{fig:anisotropies}
\end{figure}

As it can be seen in Fig.~\ref{fig:anisotropies}, the nucleation barrier in the case  $m_{2y}>m_{2x}$ is lower and the state bonded to the surface is very stable and characterized by a fractional vortex with winding only in the isotropic field. On the contrary, if the heavy direction is parallel to the normal to the surface, the barrier is higher, and the vortex in the anisotropic component is the first to enter the sample. Hence, anisotropies provide a way to enhance the stability of vortices near the boundary. This can lead to the case where fractional vortices can be stable when laying near the edges of the system aligned with one of the principal axes while being unstable near edges oriented in the orthogonal direction. This situation is shown in Fig.~\ref{fig:anisotropies}, as only one of the minimum energy paths has a clearly defined minimum which is also lower than the initial state.

\edit{
\subsection{Nucleation in type-1.5 systems}

In type-2 systems the nucleation process of consecutive vortices is not notably different from the nucleation of the first vortex, if the density of vortices is low enough. This is due to the repulsive interaction between composite vortices.

The situation is different in type-1.5 systems because of the long-range attractive interaction between vortices that tends to cluster them together. An additional nucleating vortex is, therefore, attracted by the composite vortices already present in the system. Hence, the nucleation is favored in the region of the boundary closer to the pre-existing vortex.

In Fig.~\ref{fig:type-1.5}, we compare the minimum energy path of the nucleation of the first composite vortex, with the nucleation of the second one. The nucleation barrier for the entry of a second composite vortex, in the considered example, is considerably lower than the one for the first vortex. Moreover, the presence of the attractive interaction between the vortices lowers the energy of the two-composite-vortex cluster.

Then, we consider the nucleation of a fractional vortex in the large-core component in the presence of a composite one in the bulk, i.e. $(1, 1) \to (2, 1)$. The nucleation barrier for such an event is comparable to the one for the nucleation of a second composite vortex, i.e. $(1, 1) \to (2, 2)$ . We attribute this effect to the attractive interaction in the large-core component. The resulting soliton has a $(2, 1)$ internal structure. Since it has a fractional flux, it is unstable in the bulk. However, it can be stabilized by the interaction with the boundary resulting in an unconventional surface bonded fractional soliton. Notice that this structure is more stable that the Meissner state, even if it is still metastable with respect to the two-vortex cluster.

\begin{figure}[!h]
    \centering
    \includegraphics{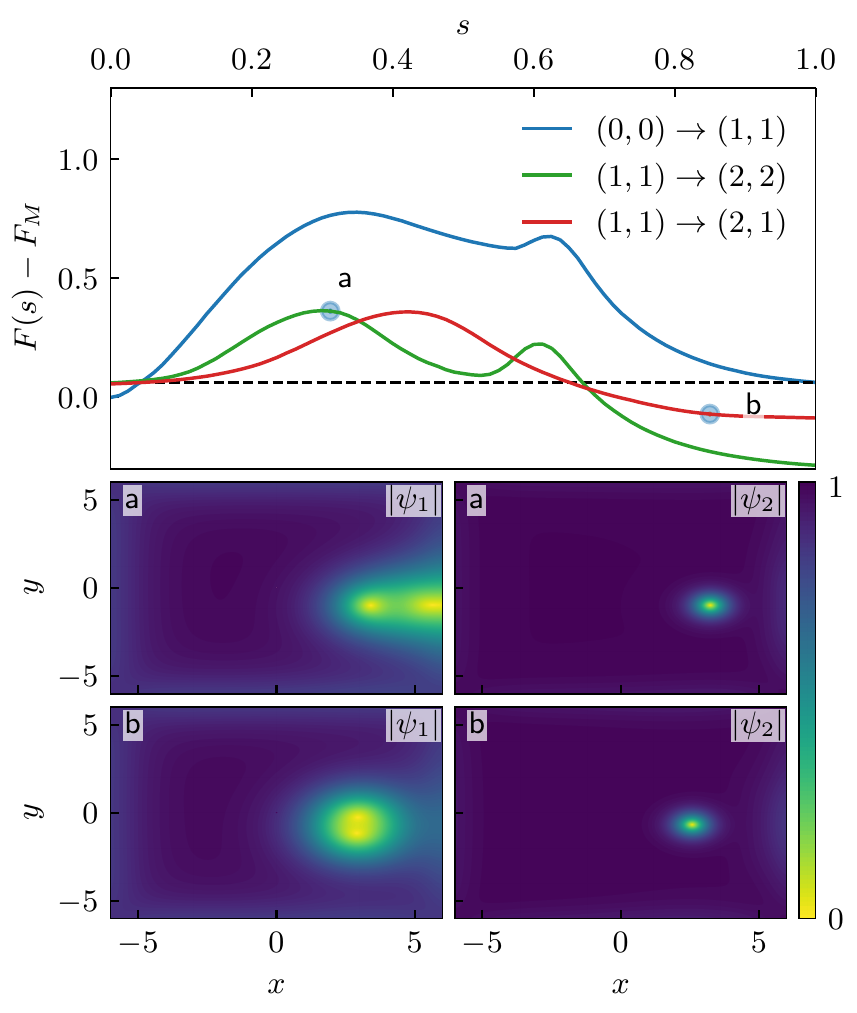}
    \caption{Minimum energy path of the first vortex nucleation compared with the nucleation of a second composite and fractional vortex in System H at $H=1.0$. The dashed black line indicates the energy of a composite vortex. The plots in the second row show the first sphaleron configuration corresponding nucleation of the $(1, 0)$ fractional vortex. In the third row, we show the configuration of the surface bonded fractional $(2,1)$ soliton.
    }
    \label{fig:type-1.5}
\end{figure}

In the fractional $(2, 1)$ soliton, the small core fractional vortex is shared between the two large-core vortices. The small energy barrier, combined with the limited increase in the system free energy for this configuration, increases the probability of seeing such solitons due to thermal fluctuations or by careful preparation of the state. This state can also be seen as containing a trapped skyrmion near the interface.

\subsection{Condensates with different charges}
All of the most common models for multicomponent superconductors assume that the charges of the condensate are equal. However, the multicomponent Ginzburg-Landau model can be extended to include systems where the charges of the two components are different but commensurate. These models describe a mixture of charged superfluids that can model more exotic systems like, e.g., the proposed coexisting electronic superconductivity with a Bose condensate of deuterons in liquid metallic deuterium \cite{babaev2004superconductor}  . In this case, unconventional vortex structures with a disparity in phase winding arise \cite{Garaud2014a, chatterjee2020chemical}.

In these systems, the composite vortices have a complex structure, due to the different phase winding required to enclose a quantum of magnetic flux. We consider, as an example, a system in which one of two condensates has a double charge with respect to the other (System D). Here, the stable soliton is a cluster with a $(2, 1)$ structure, composed of two fractional vortices in the first component, and one fractional vortex in the second component. Notice that, this structure, metastable in the conventional system, is now stable. While the $(1, 1)$ composite vortex can now live only as a metastable solution near boundaries.

We study the nucleation process of such structure in Fig~\ref{fig:disparity}, where is it possible to observe how the nucleation process is strongly fractionalized. The entrance of each fractional vortex is, in fact, associated with an energy minimum.

\begin{figure}[ht]
    \centering
    \includegraphics[width=\columnwidth]{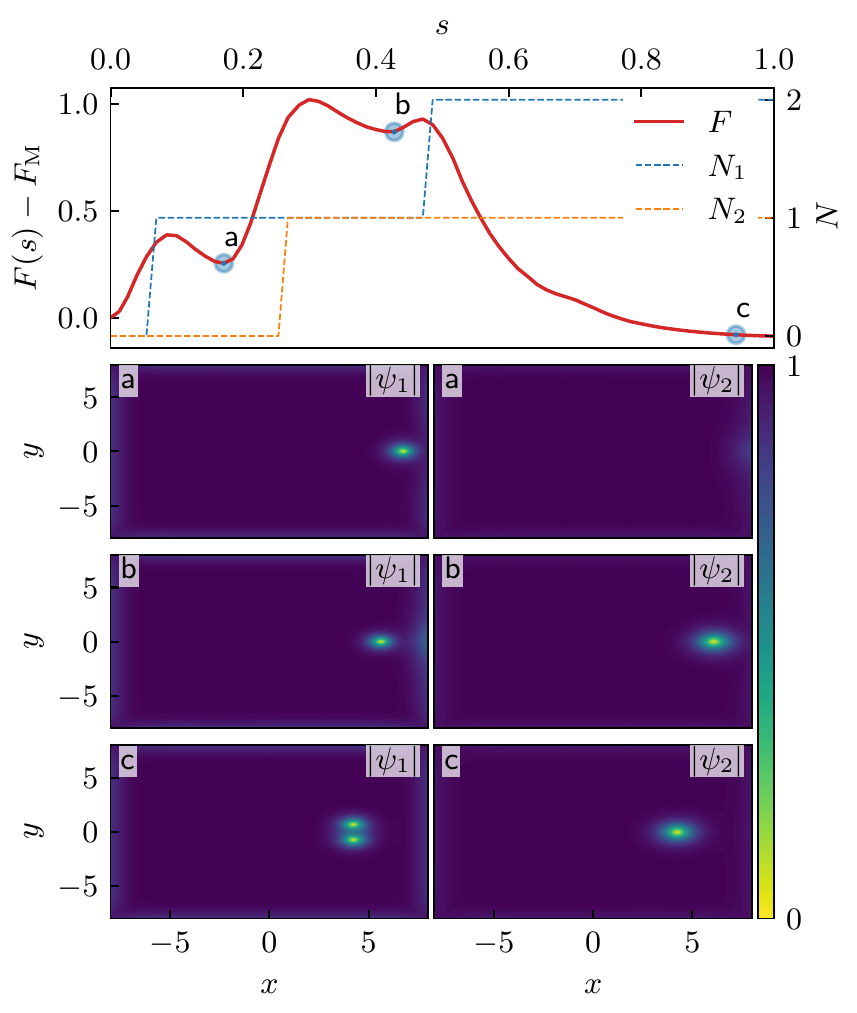}
    \caption{Minimum energy path of the nucleation of a $(2, 1)$ soliton in  System D at $H=1.2$. The top panel reports the minimum energy path of the nucleation process. The dashed lines $N_1(s)$ and $N_2(s)$ are included to help identify the nucleation of the fractional vortices. The contour plots report the system's order parameters, $\psi_1$ and $\psi_2$, in specified configurations along the minimum energy path marked by the a,b,c-points in the top panel.
    The nucleation is fractionalized in three steps with $(1, 0)$ and $(1, 1)$ as intermediate metastable states. The global minimum is achieved when $(N_1, N_2) = (2, 1)$.}
    \label{fig:disparity}
\end{figure}

Any deviation from the structure prescribed by Eq.~\eqref{eq:disparity_quantization} results in an unstable topological soliton in the bulk. However, unstable solitons can be stabilized by boundary effects, in a similar fashion to the surface bonded fractional vortices in the conventional two-component superconductors. We verified the stability of these structures by simulating the nucleation process of $(1, 0)$ and $(1, 1)$ vortices in Fig.~\ref{fig:disparity_surface}. In both cases, a sizable barrier prevents the escape of the surface bonded solitons.

\begin{figure}[ht]
    \centering
    \includegraphics[width=\columnwidth]{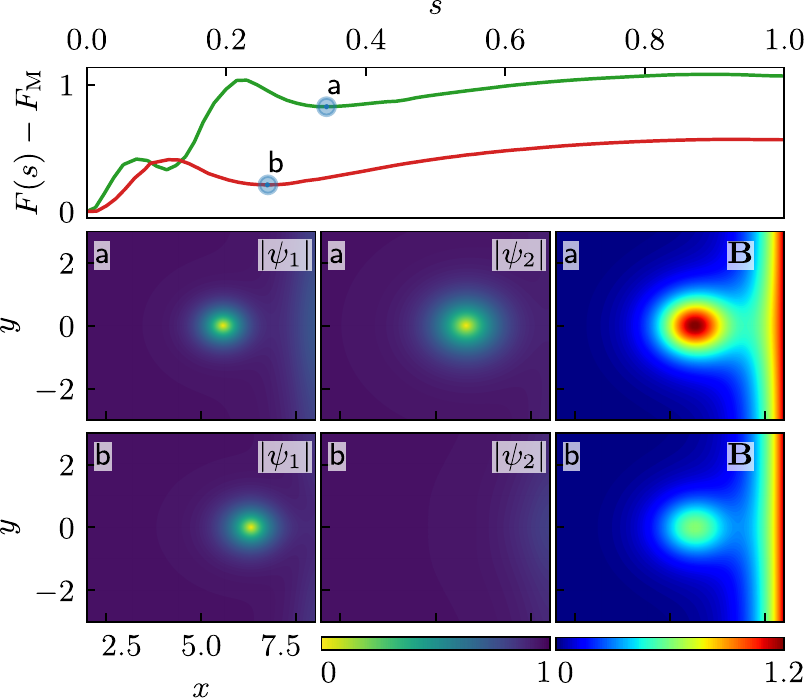}
    \caption{Surface bond  soliton in System D. Any soliton with a winding in the large core component that is smaller than the natural number (2 in this case), is unstable in the bulk and can be present only in a metastable state while bonded to the surface.
    The top panel reports the free energy for the first nucleation (green) and for the second nucleation (red) as a function of the parametrization parameter $s$. The contour plots report the system's order parameters $\psi_1$ and $\psi_2$, and magnetic field $\vb{B}$ at specified configurations in the minimum energy path marked by a blue dot in the top panel.
    \label{fig:disparity_surface}
    }
\end{figure}
}

\subsection{Effects of direct intercomponent interactions}

In this section, we consider the effect of direct interactions,  \edit{i.e. $V_\mathrm{int}\neq0$}, on the nucleation mechanism. As a start, we study the effect of bilinear Josephson on the nucleation process by simulating the nucleation of composite vortices in System E with the coupling term 
\begin{equation}
V_\mathrm{lin}\qty(\psi_1, \psi_2) = \frac{\eta}{2}\qty(\psi_1 \psi_2^* + \mathrm{c.c.}) = \eta \abs{\psi_i}\abs{\psi_2} \cos(\theta_{12}) \,.
\end{equation}
This term introduces phase-locking in the system such the spontaneously broken symmetry is reduced to $U(1)$. For positive $\eta$, the minimum is obtained for $\theta_{12}=\pm \pi$, while for negative values the minimum is for $\theta_{12}=0$. In the following, we restrict the study to positive $\eta$ as the sign can be absorbed in the phase of one field. 

\begin{figure}[h]
    \centering
    \includegraphics{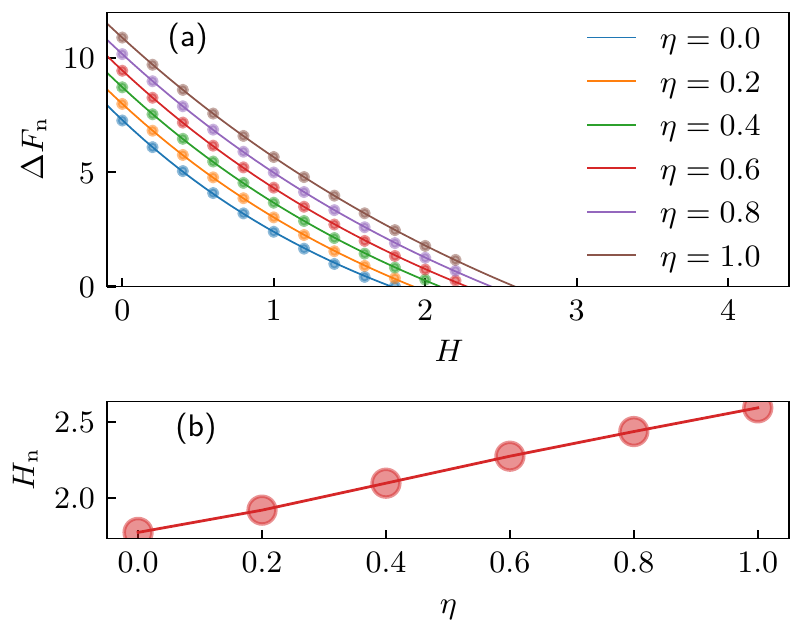}
    \caption{(a) Effect of bilinear Josephson coupling on the nucleation barrier $\Delta F_\mathrm{n}$ for composite vortex nucleation in System E with various bilinear Josephson coupling strength $\eta$ as a function of the external field $H$. The continuous lines are the estimated exponential regressions. (b) Effect of the bilinear Josephson coupling on the nucleation field $H_n$ in System E for composite vortex nucleation as a function of the bilinear Josephson coupling $\eta$ calculated by exponential fitting. The nucleation field increases approximately linearly with $\eta$.}
    \label{fig:bilinear_nucleation}
\end{figure}

As show in Fig.~\ref{fig:bilinear_nucleation}~(a), the nucleation barrier gets higher with increasing coupling. The curves can be fitted reasonably well thought an exponential regression  using $\Delta F^{\textrm{(fit)}}_\textrm{n}(H) = C_1 \qty[ 1 - \exp\qty(-C_2(H-H_\textrm{n}))]$ as the ansatz, where $C_1$, $C_2$, and $H_\textrm{n}$ are the coefficients estimated for each value of $\eta$. In this way, we estimated the nucleation field shown in Fig.~\ref{fig:bilinear_nucleation}~(b). From the simulations, it appears that, in the considered regime, the nucleation field $H_\mathrm{n}(\eta)$ increase linearly as a function of $\eta$. Note that the fractional vortex interaction potential is asymptotically linear in the presence of this coupling. 

To check the effect of bilinear Josephson coupling on fractional surface vortices, we simulated the minimum energy path of fractional vortex nucleation for System U and $H=1.2$. The bilinear Josephson coupling has a detrimental effect on surface-bond states. Already for $\eta=0.01$, the fractional surface vortex becomes a metastable state with energy higher than the Meissner state. Increasing further $\eta$, we find that, at values around $0.04$, the nucleation barrier for a vortex in the second component reaches zero such that the surface-bond fractional vortex state becomes unstable.

Let us now consider the density coupling
\begin{equation}
    V_\mathrm{den}\qty(\psi_1, \psi_2) = \frac{\gamma}{2} \abs{\psi_1}^2 \abs{\psi_2}^2 \,,
\end{equation}
this term does not break the $U(1)\cross U(1)$ gauge symmetry. Depending on the sign of $\gamma$, this term promotes an attraction or repulsion between fractional vortices in different components. A repulsive (positive $\gamma$) interaction can lead to the formation of skyrmions in the system~\cite{agterberg2014microscopic, garaud2014skyrmions}. In this case, the composite vortex formed by exactly coinciding vortex cores is a metastable soliton, while the stable skyrmion is formed by spatially separated bonded vortices.

\begin{figure}[h]
    \centering
    \includegraphics{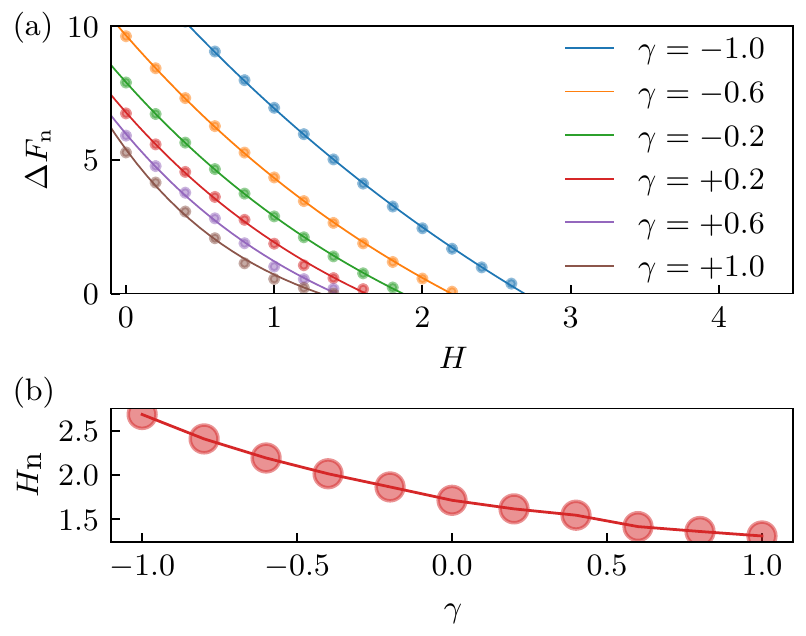}
    \caption{(a) Effect of density coupling on the nucleation barrier $\Delta F_\mathrm{n}$ of composite vortices in System E. The continuous lines are exponential fitting of the nucleation barriers. Strong biquadratic Josephson coupling changes the shape of the curve $H_\mathrm{n}(H)$ in a way that reduces the quality of an  exponential fitting.
    Panel (b) shows the composite nucleation field $H_n$ as a function of the density coupling parameter $\gamma$.}
    \label{fig:density_nucleation_fixed_gamma}
\end{figure}

We observe that the presence of repulsive density coupling decreases the nucleation barrier and fields, as shown in Fig.~\ref{fig:density_nucleation_fixed_gamma}. Since the fractionalized entry is favored, even in the case of System E, where the two fields are perfectly equal, the metastable field configuration featuring only one fractional vortex can be viewed as a partial-skyrmion state.

\begin{figure}[h]
    \centering
    \includegraphics{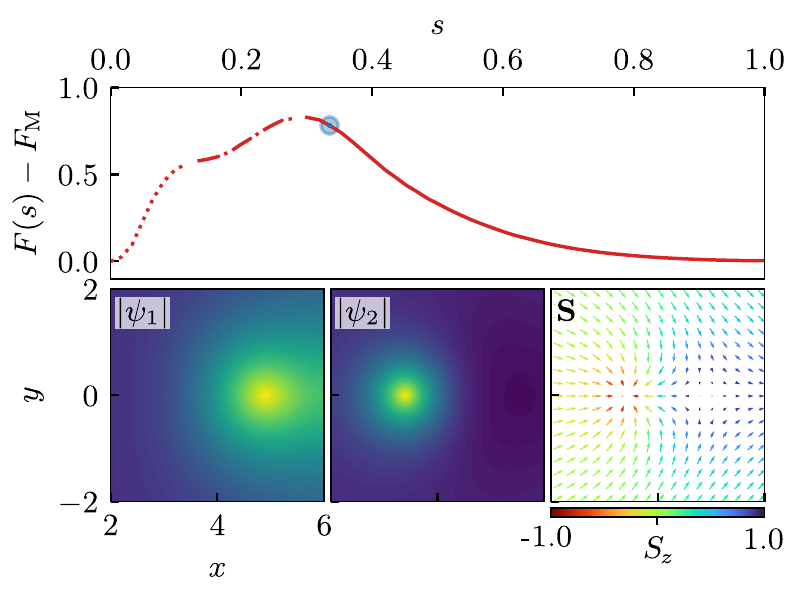}
    \caption{
    \edit{
    Nucleation process of a skyrmion in System H with $\gamma=0.3$ at $H=0.6$. The intercomponent direct interaction causes the repulsion between the two fractional vortices in the two components, favoring fractionalized entry.
    The blue dot on the top panel identifies the transition's sphaleron, whose order parameter configurations, as well as the pseudospin vector field, are displayed in the bottom panels.
    }
    \label{fig:skyrmion_entry}
    }
\end{figure}

We have tested the effect of density coupling for System U (with $H=1.2$) through a simulation of the decay of the fractional surface vortex by nucleation of a second fractional vortex in the small-core component. We observe that the nucleation barrier for this process decreases as $\gamma$ increases. Moreover, the metastable minimum state is found for a vortex further from the surface. For $\gamma\simeq0.5$ the surface vortex states spontaneously decay and multiple vortices in the small-core component nucleate. 
\edit{An example of a skyrmion entry process is shown in Fig.~\ref{fig:skyrmion_entry}, for system H with a magnetic field close to $H_{\rm c1}$. Nonetheless, for some type-2 systems, the partial-skyrmion can still be metastable even in the presence of a positive density coupling.}

Negative $\gamma$ has also a detrimental effect on surface vortex states, but with a different mechanism. The direct interaction leads to a reduction of the escape barrier for the fractional vortex such that, already for $\gamma=-0.2$, we observe spontaneous decay to a Meissner state. 

\section{Conclusions}

In this paper, we studied the minimum energy path of the vortex nucleation process for the Ginzburg-Landau model of two-component superconductors. We obtained solutions that feature multiple sphalerons connected to the entry of a single fractional vortex. Even composite vortices enter the sample through fractionalized nucleation in many systems. 
In the type-2 and type-1.5 regime, in agreement with the previous studies based on the London-model~\cite{silaev2011stable}, fractional vortices with winding in the large-core component can be stabilized by the interaction with boundary currents. We characterized the stability of these solutions and the conditions for their appearance in a fully nonlinear theory.

\edit{Even when surface fractional vortices do not appear in the ground state of the system, but only as metastable solutions, this does not preclude the observability of such states, as their lifetime is, in general, not zero. Moreover, these metastable states can be stabilized by additional effects. We found, in particular, that mass anisotropy provides the strongest stabilization mechanism.}

The method that we developed allowed for the inclusion and analysis of the effect of direct interaction between the fields like the bilinear Josephson coupling and density coupling. We find that fractional vortices are metastable near boundaries even for nonzero Josephson coupling. Note that, in some of the considered cases, the barrier for the nucleation process of the vortex in the second component is very low. For this reason, we do not expect the surface vortex states to be generically observed if not actively searched in in well-controlled experiments. \edit{In type-1.5 superconductors, vortex clusters with a total flux that is not an integer multiple of the flux quantum may appear as metastable solitons near boundaries.} 

These surface-trapped fractional objects can be observed in scanning probes and can serve as smoking-gun evidence of multicomponent order parameters in multiband superconductors, superconductors that break multiple symmetries, including models proposed for superconductivity in magic-angle twisted bilayer graphene~\cite{khalaf2020symmetry,chichinadze2020nematic}. Applied currents can move these solitons along boundaries of samples with a high-quality surface, making them distinguishable from ordinary vortices pinned to a defect.

The formation of solitons with fractional magnetic flux can also be affected by the surface effects discussed in~\cite{samoilenka2020bcs}. This opens up the possibility of the creation of a new platform for fluxonic information processing  \cite{golod2015single}.

As a secondary result, we have shown how the gauged string method can be applied to a system with additional gauge symmetries, like the $U(1)\times U(1)$ model, confirming the flexibility of this method. This can be easily extended to systems with other gauge symmetries or used to study other processes like vortex clustering in type-1.5 systems. 

\section{Acknowledgments}
We gratefully acknowledge the support of NVIDIA Corporation with the donation of the Quadro P6000 GPU used for this research. A.M. was supported by the Danish National Research Foundation, the Danish Council for Independent Research \textbar Natural Sciences. A.B. and E.B. were supported by the Swedish Research Council Grants No. 2016-06122, 2018-03659, and G\"{o}ran Gustafsson Foundation for Research in Natural Sciences.

\bibliography{bibliography.bib}

\end{document}